\documentclass{aastex}
\usepackage{emulateapj5}
\usepackage{onecolfloat5}
\usepackage{apjfonts}
\usepackage{epsfig}

\def\gsim{\;\rlap{\lower 2.5pt
 \hbox{$\sim$}}\raise 1.5pt\hbox{$>$}\;}
\def\lsim{\;\rlap{\lower 2.5pt
   \hbox{$\sim$}}\raise 1.5pt\hbox{$<$}\;}
\newcommand{\be}{\begin{equation}}
\newcommand{\beq}{\begin{equation}}
\newcommand{\ba}{\begin{eqnarray}}
\newcommand{\ee}{\end{equation}}
\newcommand{\eeq}{\end{equation}}
\newcommand{\ea}{\end{eqnarray}}

\newcommand{\msun}{$M_{\odot}\hspace{1mm}$}
\newcommand{\wmap}{{\it WMAP }}

\newcommand{\unitE}{$\rm{ergs}\hspace{1mm}\rm{ cm}^{-2\hspace{1mm}}\rm{ s}^{-1}\hspace{1mm}\rm{ deg}^{-2}$\hspace{1mm}}
\newcommand{\hs}{\hspace{1mm}}
\usepackage{natbib}

\begin{document}
\twocolumn[


\title{A Limit from the X-ray Background on the Contribution of Quasars to Reionization}

\author{Mark Dijkstra, Zolt\'an Haiman}

\affil{Department of Astronomy, Columbia University, 550 West 120th Street, New York, NY 10027}

\author{Abraham Loeb}

\affil{Harvard-Smithsonian Center for Astrophysics, 60 Garden Street, Cambridge, MA 02138}
\vspace{0.23cm}
\vspace{-0.5\baselineskip}
 
\begin{abstract}
A population of black holes (BHs) at high redshifts ($z\gsim 6$) that
contributes significantly to the ionization of the intergalactic
medium (IGM) would be accompanied by the copious production of hard
($\gsim 10$ keV) X-ray photons.  The resulting hard X--ray background
would redshift and be observed as a present--day soft X--ray
background (SXB).  Under the hypothesis that BHs are the main
producers of reionizing photons in the high--redshift universe, we
calculate their contribution to the present--day SXB.  Our results,
when compared to the unresolved component of the SXB in the range
0.5-2 keV, suggest that accreting BHs (be it luminous quasars or their
lower--mass ``miniquasar'' counterparts) did not dominate
reionization.  Distant miniquasars that produce enough X--rays to only
partially ionize the IGM to a level of at most $x_e\sim 50\%$ are
still allowed, but could be severely constrained by improved
determinations of the unresolved component of the SXB.
\end{abstract}

]



\section{Introduction}
\label{sec:intro}
The recent discovery of the Gunn--Peterson (GP) troughs in the spectra
of $z>6$ quasars in the Sloan Digital Sky Survey (SDSS; White et
al. 2003; Wyithe \& Loeb 2004) suggests that the end of the
reionization process occurs at a redshift near $z\sim 6$.  At this
epoch, the ionizing sources drive a strong evolution of the neutral
fraction of the intergalactic medium (IGM) from values near unity down
to $x_{\rm HI}\sim 10^{-3}$ (e.g., Fan et al. 2002).  On the other
hand, the high electron scattering optical depth, $\tau_e=0.17 \pm
0.04$, measured recently by the {\it Wilkinson Microwave Anisotropy
Probe (WMAP)} experiment (Spergel et al. 2003) suggests that ionizing
sources were abundant at a much higher redshift, $z\sim 15$.  These
data imply that the reionization process is extended and complex, and
is probably driven by more than one population of ionizing sources (see,
e.g., Haiman 2003 for a post-{\it WMAP} review).

The exact nature of these ionizing sources remains unknown.  Natural
candidates to account for the onset of reionization at $z\sim 15$ are
massive, metal--free stars that form in the shallow potential wells of the
first collapsed dark matter halos (Wyithe \& Loeb 2003a; Cen 2003a; Haiman
\& Holder 2003).  The completion of reionization at $z\sim 6$ could then be
accounted for by a normal population of less massive stars that form from
the metal--enriched gas in larger dark matter halos present at $z\sim 6$.

The most natural alternative cause for reionization is the ionizing
radiation produced by gas accretion onto an early population of black
holes (``miniquasars''; see Haiman \& Loeb 1998, Wyithe \& Loeb 2003c,
Bromm \& Loeb 2003).  The ionizing emissivity of the known population
of quasars diminishes rapidly beyond $z\gsim 3$, and bright quasars
are unlikely to contribute significantly to the ionizing background at
$z\gsim 5$ (Shapiro, Giroux \& Babul 1994; Haiman, Abel \& Madau 2001;
Wyithe \& Loeb 2003a).  However, if low--luminosity, yet undetected
miniquasars are present in large numbers, they could dominate the
total ionizing background at $z\sim 6$ (Haiman \& Loeb 1998). Recent
work, motivated by the {\it WMAP} results, has emphasized the
potential significant contribution to the ionizing background at the
earliest epochs ($z\sim 15$) from accretion onto the seeds of
would--be supermassive black holes (Madau et al. 2003; Ricotti \&
Ostriker 2003).  The soft X--rays emitted by these sources can
partially ionize the IGM early on (Oh 2001; Venkatesan \& Shull 2001).

A population of miniquasars at $z\gsim 6$ would be accompanied by the
presence of an early X-ray background.  Since the IGM is optically
thick to photons with energies $E$ below $E_{\rm max} = 1.8
[(1+z)/15)]^{0.5} x_{\rm HI}^{1/3}\hs {\rm keV}$, 
the soft X-rays with $E\lsim E_{\rm max}$ would be consumed by neutral
hydrogen atoms and contribute to reionization.  However, the background of
harder X-rays would redshift without absorption and would be observed as a
present--day soft X--ray background (SXB).  In this paper, {\it we examine
the hypothesis that accreting BHs are the main producers of reionizing
photons in the high--redshift universe, and calculate their contribution to
the present--day SXB in this case}.  Our results, when compared to the
unresolved component of the SXB, suggest that accreting BHs cannot contribute
significantly either to the completion of reionization at $z > 6$, or to
the partial ionization of the IGM at $z\gsim 15$ to ionized fractions of
$x_e\gsim 0.5$.

The outline of the paper is as follows:
In \S~\ref{sec:method}, we describe the method to calculate the SXB
from quasars that contribute to reionizing the universe.
In \S~\ref{sec:sxrb}, we critically discuss current X-ray
observations, focusing on the unresolved fraction of the SXB that
could be attributed to distant quasars.
In \S~\ref{sec:fullion}, we calculate the expected contribution to the
SXB from hypothetical quasars and their lower--mass miniquasar counterparts
that fully ionized the universe.
In \S~\ref{sec:preion}, we repeat our analysis for a putative miniquasar
population that partially ionizes the IGM at high redshifts.
In \S~\ref{sec:discuss}, we discuss how various simplifications
made in our analysis influence our final results.
Finally, in \S\ref{sec:conclusions} we summarize our results and the
implications of this work.
Throughout this paper, we adopt the background cosmological parameters
as measured by the {\it WMAP} experiment, $\Omega_m=0.27$,
$\Omega_{\Lambda}=0.73$, $\Omega_b=0.044$, and $h=0.71$ (Spergel et
al. 2003) and set the mass fraction of helium to $Y_{\rm He}=0.24$
( e.g. Burles, Nollett, \& Turner, 2001).
In the rest of the paper `accreting BHs' will refer
to both quasars and their lower--mass ``miniquasar'' counterparts.

\newpage
\section{Modeling the Contribution to the SXB}

\label{sec:method}

To calculate the contribution to the present--day SXB from accreting BHs that
reionized the universe, we assume for simplicity that the accreting BHs form in a
sudden burst at redshift $z=z_Q$.  The spectrum of the ionizing background
is a crucial ingredient of the modeling, and depends on the type of accreting
BH that is considered. The details of the assumed emission spectrum 
for each type are discussed in \S\ref{sec:fullion} and \ref{sec:preion}.
In \S~\ref{sec:discuss}, we examine the dependence of
our results on the emission spectrum.

Assuming that the cumulative hard X--ray flux of the accreting BHs results
in a present--day X-ray background $F_E$, we can compute the number of
H--ionizing photons present at redshift $z_Q$ per unit comoving
volume:
\begin{eqnarray}
n_{\gamma}=\frac{4\pi}{c} \int_{\frac{13.6}{1+z_Q} \rm
{eV}}^{\frac{E_{\rm{max}}}{1+z_Q} \rm {eV}} dE \frac{F_E}{E}
\label{eq:ngamma}
\end{eqnarray}
In the above equation, $E$ and $F_E$ correspond to photon energies and
intensities at $z=0$, and $F_E$ has the units of ${\rm
ergs~cm^{-2}~s^{-1}}$ $\rm{~deg^{-2}~erg^{-1}}$.  The photon energy
$E_{\rm max} = 1.8 [(1+z)/15)]^{0.5} \hs {\rm keV}$ reflects the value
above which the IGM becomes optically thin across a Hubble length for
a fully neutral IGM\footnote{More precisely, the
upper limit should be taken as the energy below which photons would
typically be absorbed and contribute to H-reionization at any redshift
prior to $z\approx 6$ for a partially ionized IGM --
rather than $E_{\rm max}$ as defined above (see \S\ref{sec:preion}, eq. \ref{eq:emax}). 
In practice, the difference between the two definitions is small.}.  
For most spectral slopes in the range we study
here, we have verified that our results are insensitive to the choice
of $E_{\rm max}$.  The exception is in \S\ref{sec:preion}, where we
consider unusually hard spectra; in this case the exact value of
$E_{\rm max}$ does influence the results. A brief discussion of this will
be given in \S\ref{sec:preion}.

The total number of H--ionizations per unit comoving volume is
approximately equal to the comoving density of H--ionizing photons $n_{\gamma}$
 with energy below $\approx E_{\rm max}$. This simple formulation
ignores secondary ionizations by the photo--electrons produced by the more
energetic photons ($E \gtrsim 100$ eV).  For spectra of the form $F_E
\propto E^{-\alpha}$ with $\alpha > 0.7$ at $E > 13.6$ eV, these secondary
ionizations are subdominant compared to the ionizations by the UV--photons
(Abel et al. 1997).  More importantly, the number of
secondary ionizations is negligible for any spectrum if the ionized medium
has a neutral fraction, $x_{\rm HI} \ll 1$ (since the electron energy is instead
used to heat the gas; e.g. Shull \& van Steenberg, 1985).
For harder spectra and a partially ionized medium, secondary ionizations
become important.  We will return to this issue in \S ~\ref{sec:preion}
below, where we discuss a miniquasar population that is assumed to only
partially ionize the IGM.

Since helium may be reionized later by the already known population of
optical quasars at $z\sim 3$ (Sokasian, Abel \& Hernquist 2002; Wyithe
\& Loeb 2003a), we consider only hydrogen here.  The presence of
helium will not significantly change our results, as will be discussed
in \S\ref{sec:discuss:helium} below.

The normalization of the present--day background $F_E$ is obtained by
requiring that the quasar population would lead to reionization,
\be
n_{\gamma}=\eta n_{H,0}
\label{eq:eta}
\ee where $n_{H,0}=\Omega_bh^2 \rho_{\rm crit}(1-Y_{\rm He})/m_p=2.05
\times 10^{-7} \hs \rm{cm}^{-3}$ is the number density of H--atoms at
$z=0$ and $\eta $ is the number of ionizations per
hydrogen atom that are required to achieve reionization. 

The value of $\eta$ is larger than unity because of the need to balance
recombinations. The ratio between the Hubble time, which is taken
 to be $t_{\rm hub}(z)\equiv 2H^{-1}(z)/3$, and the recombination
time in a region at overdensity $\delta$ that is partially ionized to
$x_{H^{+}}\equiv n_{H^{+}}/n_H$, where $n_H$ is the number density of hydrogen
atoms (neutral and ionized) at any redshift $z$, is given by:

\be 
\frac{2H^{-1}}{3t_{\rm rec}}= 1.7\hs C \hs \Big{(}
\frac{10^4}{T}\Big{)}^{0.7}\Big{(} \frac{1+z}{11}\Big{)}^{1.5}x_e\hs(1+\delta)
\label{eq:recomb1}
\ee
Where $x_e\equiv n_e/n_H$ ($=x_{H^{+}}$, because helium is not included),
$C\equiv \langle n^2 \rangle/\langle n \rangle ^2$ which depends on
the level of clumping in the high--redshift the IGM, and is expected
to be $\gtrsim 10$ (Haiman, Abel, \& Madau 2001; Wyithe \& Loeb 2003a;
Ricotti \& Ostriker 2003), and $T$ is the gas temperature. This
implies that recombinations cannot be ignored at high redshift. In
fact, the number of recombinations per hydrogen atom between $z=z_Q$
and $z=7$, $N_{\rm rec}$, can be estimated from

\begin{equation}
N_{\rm rec}\hs\equiv\hs \int_{t(z=z_Q)}^{t(z=7)}\frac{\alpha_H(T)\hs n_e \hs n_{H^{+}}}{n_{H^{+}}}dt
\label{eq:recomb2}
\end{equation}

Interestingly, for any combination of $z_Q$ and $x_e$ yielding the same $\tau_e$,
this can be written as a function of $\tau_e$
as long as $x_e$ is constant with redshift for $z > 7$ and $x_e \sim 1$ for $0 <z <7$:
\be
N_{\rm rec} = 2.4\hs C\hs (1+\delta)\Big{(}
\frac{T}{10^4}\Big{)}^{-0.7}\Big{(} \frac{\tau_e-0.05}{0.12}\Big{)},
\label{eq:recomb3}
\ee
where $\tau_e$ is the electron scattering optical depth between redshifts
$z=0$ and $z=z_Q$.  Note that $x_e \sim 1$ for $0 <z <7$ and $x_e=0$ for
$z>7$ yields $\tau_e=0.05$.  The recombination coefficient $\alpha_H(T)$ is
taken from Abel et al. (1997). To maintain a fixed
ionization fraction we need $\eta \equiv N_{\rm rec}+1 \sim 25$ ionizations per hydrogen
atom. In \S\ref{sec:eta} we will show that the actual value of $\eta$
depends on the exact topology of reionization, and therefore quite
uncertain. The SXB intensity $F_E$ predicted below simply scales linearly
with the exact value of $\eta$. Given the uncertainty in $\eta$ we will
adopt the fiducial value of $\eta=10$, and scale our quoted fluxes in units
of $(\eta/10)$ (see Table 2).

It is important to emphasize that we are normalizing $F_E$ based on
the total ionizing background, required to produce reionization. Our
prediction for $F_E$ therefore does not require the knowledge of the
luminosity function of the quasars.

\vspace{3\baselineskip}
\section{The Unresolved SXB and Its Uncertainties}
\label{sec:sxrb}

Since the unresolved fraction of the SXB is the key observable used
to obtain the constraints in this paper, we first discuss its measurement
and uncertainty.  The soft X-ray band in the energy range $0.5-2.0$ keV has
been studied by Moretti et al. (2003, hereafter M03).  We focus on this
energy range since it provides the strongest constraints. The $0.5-2$ keV
band corresponds to hard X--rays at $z\geq 6$, which are well within the
regime of photon energies to which the IGM is optically thin.  M03
determines the intensity of the total SXB within this band to be $ \int F_E
dE \sim EF_E(1 \hs \rm{keV})$ = $7.53 \pm 0.35 \times 10^{-12}$ \unitE,
when combining 10 different measurements reported in the literature.  M03
further include deep pencil beam surveys together with wide field shallow
surveys and ultimately are able to identify discrete sources in the flux
range $2.44 \times 10^{-17}-1.00 \times 10^{-11}$
$\rm{ergs}\hspace{1mm}\rm{ cm}^{-2\hspace{1mm}}\rm{ s}^{-1}$\hspace{1mm}
within the soft band.  They find that $94^{+6}_{-7} \%$ of the SXB is made up
of discrete X-ray sources. The majority of this resolved component involves
point sources ($88 \pm 7 \%$ of the total SXB), with extended sources
amounting to $6\%$ of the background. Barger et al. (2002, 2003) show that
the bulk of these sources are at low redshifts, $z < 4$.  In
\S~\ref{sec:discuss:point} below we find that any $z \geq 6$ quasars that
``hide'' among the already resolved point sources can only make up to $1\%$ of
the total SXB.

The fraction of the SXB contributed by known discrete (point or extended)
sources depends on the sensitivity limits of current observations. M03 find
that when they extrapolate the analytic form of the log N-log S
distribution (cumulative fraction of the total SXB in sources brighter than
flux S) to $S_{\rm min}\sim 3 \times 10^{-18}~{\rm ergs~s^{-1}~cm^{-2}}$,
which is a factor of $\sim$ 10 lower than current sensitivity limits, the
{\it entire} SXB can be accounted for by discrete sources. The additional,
still undetected faint X--rays sources could be either point--like or
extended.

Wu \& Xue (2001) calculate the theoretically expected amount of X--ray
emission in the soft band from thermal emission by gas in all clusters and
groups in the universe using the observed $L_X-T$ relation and X--ray
luminosity function (XLF) and find this should be $1.18 \times 10^{-12}$
\unitE. This is $16\%$ of the total SXB quoted by M03. Approximately $\sim
40\%$ of this contribution originates in galaxy groups; ignoring groups
(where the XLF extrapolation becomes uncertain) provides a conservative
estimate of the total expected contribution from thermal emission from extended sources, 
$\sim 0.6 \times 1.18 \times 10^{-12}=0.68\times 10^{-12}$ \unitE, which is
$9\%$ of the total SXB. This suggests that the actual contribution to the
SXB from extended sources is higher than claimed by M03, and that
deeper X--ray observations will reveal new extended X--ray sources at flux
levels below current sensitivity limits.  To remain conservative we take
the M03's estimate of $6\%$ for the contribution of extended sources to the
SXB, but keep in mind that the actual number is probably higher, leaving a
smaller residual that can be attributed to distant accreting BHs.

Recent numerical simulations by Dav{\' e} et al. (2001) suggest that a
fraction as large as 0.3-0.4 of all baryons might be in the warm/hot
intergalactic medium (WHIM) with temperatures between $10^5-10^7$K,
residing in diffuse large scale structures such as the filaments
connecting virialized objects at $z=0$. The WHIM should emit some
thermal radiation in the soft X--ray band, but has so far not been
detected. In this paper, we ignore the possible contribution to the
SXB from the WHIM. Note that any WHIM contribution would strengthen
our results, since it would again leave a smaller residual of the SXB
that can be attributed to distant accreting BHs.

Soltan (2003) pointed out that point sources can never make up the
full SXB, because the X-ray photons will be Thompson scattered by free
electrons, yielding a diffuse component. This component is estimated
to be $\sim 1.0-1.7\%$ of the total SXB.  Although this is a small
fraction, it makes a non-negligible contribution to the unresolved
SXB, and we include it in our analysis below.

A summary of the possible SXB contributions is given in
Table~\ref{table}.  The Table quotes an intensity for a ``mean'' and a
``maximum'' unaccounted SXB flux.  The mean unaccounted flux is the
total flux minus the contributions from point sources, galaxy
clusters, and Thompson scattered X-rays. The maximum unaccounted flux
is calculated by summing the $1\sigma$ errors ($3^{{\rm rd}}$ column
in Table~\ref{table}) in the most conservative way: the total SXB is
set to its maximum ($+1\sigma$) value, while all other components
(point sources, extended sources and scattered X--rays) subtracted
from this total are assumed to be at their minimum ($-1\sigma$)
levels.  We find the mean and maximum unaccounted fluxes to be $0.35 \times
10^{-12}$ \unitE and $1.23 \times 10^{-12}$ \unitE, respectively.

It should be noted that a similar analysis can be done for the hard
X--Ray background (2.0-8.0 keV), but since the constraints in this
band are considerably weaker, we do not discuss them here.

\begin{table}[ht]\small
\caption{Contributions to the Soft X-ray Background}
\label{table}
\begin{center}
\begin{tabular}{lcc}
\tablewidth{3in}
Source                   & Flux   & $\sigma_{\rm flux}$\\
 & ($10^{-12}$ \unitE)   & \\
\hline
\hline
Total SXB\tablenotemark{1}                 &  7.53       & 0.35   \\
Point Sources\tablenotemark{1}             &  6.63       & 0.35   \\
Diffuse Component\tablenotemark{1}            &  0.45       & 0.25   \\
Expected scattered background\tablenotemark{2}&  0.10       & 0.03   \\ 
Unaccounted flux (mean)  &  0.35       & 0.60   \\
Unaccounted flux (max)   &  1.23       &      \\
\hline
\end{tabular}\\[12pt]
\end{center}
\tablenotetext{1}{From Moretti et al. 2003.}
\tablenotetext{2}{Diffuse Thompson--scattered radiation from point sources (Soltan 2003) }
\end{table}

\section{Full Reionization by the UV Emission of Accreting Black Holes}
\label{sec:fullion}
\subsection{Reionization by Quasars}
\label{sec:quasars}

We follow the procedure outlined in \S\ref{sec:method} and calculate
the present day SXB, $F_E$, for a burst of quasars at redshifts
$z_Q=6, 10$ and 20, assumed to fully reionize the universe.  
We adopt the shape of the fiducial quasar spectrum
from Sazonov, Ostriker, \& Sunyaev (2004, see their eqs. 8 and 14), which
is then assumed to redshift to $z=0$ without absorption for hard
X-rays. This broad--band spectral energy distribution of the average quasar
in the universe is computed based on a detailed theoretical model that fits
a range of data, including published composite spectra of quasars in the
optical, UV and X--ray bands.  The template spectrum scales
approximately as $F_E\propto E^{-1.7}$ for $E > 13.6$ eV, but becomes much
shallower beyond $E=2$ keV, where it scales as $F_E\propto E^{-0.25}$.
The main reason for this 'kink' in the spectrum around $E=2$ keV is to
accommodate the rise in the observed cosmic X--ray background in the energy
range 1-30 keV.

\begin{figure}[ht]
\vbox{ \centerline{\epsfig{file=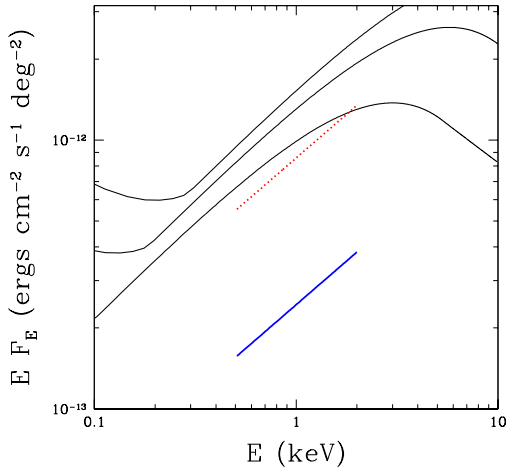,width=8.0truecm}}
\caption{Predicted soft X-ray background due to a burst of quasars
producing $\eta=10$ ionizing photons per hydrogen atom and reionizing
the universe at $z=6,10$ and $20$ (first three curves from the top).
The template spectrum for these quasars is taken from Sazonov,
Ostriker, \& Sunyaev (2004). The two straight lines plotted in the
range 0.5-2.0 keV are the maximum and mean unaccounted flux in the
observed soft X-ray background (Table~\ref{table}). A power-law
spectrum is assigned to this flux, with a slope matching the spectral
slope of quasars.}
\label{fig:saz}}
\vspace*{0.5cm}
\end{figure}

The resulting flux density is plotted in the range 0.1--10 keV in
Figure \ref{fig:saz}, assuming $\eta=10$, with the upper curve
corresponding to $z_Q=6$, and the lower curve to $z_Q=20$.  The $EF_E$
curves are redshifted and suppressed by a factor of $(1+z_Q)$ relative
to the rest--frame template spectrum (which peaks at $E=60$ keV).  As
a result, we obtain the largest SXB for the quasar burst at the low
end of the redshift range, $z_Q=6$.

The two straight lines plotted in the range 0.5-2.0 keV denote the
mean and maximum unaccounted flux of Table~\ref{table}.  The
unaccounted flux is assigned a power-law spectrum with a slope
matching the spectral slope of quasars in the range $0.5-2.0$ keV.
The figure shows that at all three redshifts, the SXB intensity
significantly exceeds the level of the mean unaccounted flux (by a
factor of $\sim 4-6$), and is above our inferred maximum level.  To
compare directly with the values quoted in Table~\ref{table} we
calculate $F_{\rm QSO} \equiv \int_{0.5 \rm{keV}}^{\rm{2.0 keV}} F_E
dE$.  For $z=6, 10,$ and $20$, we find $F_{\rm QSO}=2.17, 1.84,$ and
$1.37 \hs $ $\times 10^{-12}(\eta/10)$ ${\rm
ergs~cm^{-2}~s^{-1}~deg^{-2}}$, respectively. For any redshift, the
maximum unaccounted flux is saturated.

\subsection{Reionization by Miniquasars}
\label{sec:miniquasars}

In the previous section, we addressed the scenario in which the IGM is
fully ionized by luminous quasars.  An alternative scenario is that in
which the ionizing background is dominated by accretion onto the much
less massive seed BHs that later grow to be the massive
BHs powering luminous quasars.  The main difference, for our purposes,
is the typical spectrum of the ionizing sources in the two
scenarios. Our fiducial template from Sazonov, Ostriker, \& Sunyaev
(2004) that we used above is appropriate for luminous quasars at lower
redshifts ($0< z < 5$), powered by supermassive black holes with
masses of $M_{\rm bh}\approx 10^{6-9}$ \msun. The spectra of
miniquasars, with BHs whose masses are in the range $M_{\rm bh}\approx
10^{2-4}$ \msun (Madau et al, 2003) are likely to be harder.

To obtain the spectrum of a typical miniquasar, we follow the approach
of Madau et al. (2003) and adopt a template that consists of two
components. One component originates from a multi color accretion disk
(MCD), which describes the accretion disk as series of blackbody
annuli with different temperatures, and results in a spectrum that
scales as $F_{E,{\rm MCD}} \propto E^{1/3}$ up to $kT_{\rm max}=1\rm
{keV}\hs (M_{\rm bh}/M_\odot)^{-1/4}$.  The second component is a
simple power-law $F_{E,{\rm PL}} \propto E^{-1}$ for $E > 13.6$
eV. The precise origin of this power--law emission is not fully
understood, but is likely to be a combination of Bremsstrahlung,
synchrotron, and inverse Compton emission by a non-thermal population
of electrons.

This power-law + MCD composite model adequately describes some
ultraluminous X-ray sources (ULXs), that are claimed to be associated
with an intermediate mass black hole of $\sim 10^3$ \msun (Miller et
al. 2003). In these spectra the flux from the MCD, $F_{\rm MCD}$ , is
comparable to the flux in the power--law component, $F_{\rm PL}$.
Miller et al. (2003) find the ratio of $F_{\rm PL}$ to the total flux
to be 0.33 and 0.63 for the two ULX sources NGC 1313 X--1 and X--2
respectively.  Motivated by these observations, here we take $F_{\rm
MCD}\equiv\int_{0.0136~{\rm keV}}^{kT_{\rm max}}F_{E,{\rm MCD}}dE$,
$F_{\rm PL}\equiv\int_{0.2~{\rm keV}}^{10~{\rm keV}}F_{E,{\rm PL}}dE$,
and define the ratio $\Psi\equiv F_{\rm MCD}/F_{\rm PL}$.  
In this paper, we adopt $\Psi\equiv 1$, and vary it between 0.5 and 2.0 as an
additional source of uncertainty\footnote{ The definition of $F_{\rm PL}$
is the same as in Miller et al. (2003), whereas our definition of $F_{\rm
MCD}$ is not exactly the same because our expression for $F_{E,{\rm MCD}}$
is only an approximation to the actual spectrum (see Fig 3. in Shakura \&
Sunyaev 1973). The error this introduces is well within the uncertainties
we assign to the value of $\Psi$. }.

An additional source of uncertainty is the mass of the typical black
hole, $M_{\rm bh}$. Madau et al. (2003) show the mass function of
accreting intermediate mass black holes at four different redshifts.
The median mass at $z=22$ is $M_{\rm bh} \sim 100 \hs M_{\odot}$,
increasing to $\sim 1000 \hs M_{\odot}$ at $z=13$.  To approximately
mimic this increase, we adopt $\log(M_{\rm bh})=(4.1-0.1z) \pm
0.5$. The error in the black hole mass is incorporated into the error
in the final contribution to the SXB.

Applying the analysis outlined in \S~\ref{sec:method} with this modified
spectral shape yields $F_{\rm mQSO} \equiv \int_{0.5 \rm{keV}}^{\rm{2.0
keV}} F_E dE$ = $3.3 \pm 1.1$, $2.3 \pm 0.8$ and $1.4 \pm 0.4 \hs $ $\times
10^{-12}(\eta/10)$~\unitE for $z=6, 10$ and $20$, respectively.  The
dominant contribution to our quoted error is the uncertainty in $\Psi$, the
relative importance of the flux in the power law component to the flux from
the MCD component.  Despite the differences in the assumed spectra for
quasars and miniquasars, their contribution to the current SXB, under the
assumption that they fully ionized the IGM, is comparable.  For any
redshift, we find that miniquasars saturate the mean unaccounted flux of
the SXB. Only for $z =20$ the maximum unaccounted flux of the SXB is not
saturated within the uncertainty of our model.

\section{Partial ``Preionization'' by X--rays from Miniquasars}
\label{sec:preion}

We have so far omitted secondary ionizations from our analysis.  Secondary
ionizations are collisional ionizations by the photo--electron that is
knocked off the hydrogen atoms when ionized by energetic X--ray
photons. Shull \& van Steenberg (1985) quantify the fraction
$\phi(x_{H^{+}},E)$ of the energy of these electrons that is used for
H--ionizations.  They find that for energies $\gg 100$ eV, $\phi(x_{H^{+}},E)$
is independent of energy, and $\phi(x_{H^{+}},E) \sim 1/3$ for a neutral
medium ($x_{H^{+}}=0$) although it drops rapidly for $x_{H^{+}} \gtrsim 0.1$.
This justifies for our statement in \S~\ref{sec:method} that secondary
ionizations do not contribute significantly to full reionization (as we
further show in more detail at the end of this section).  However, secondary
ionizations can become significant in a partially ionized medium,
especially for harder spectra.

The large electron scattering optical depth \wmap result can be explained
with a {\em partially} ionized IGM, followed by full reionization at $z\la
6.3$ (e.g. Wyithe \& Loeb 2003b, 2004; Cen 2003b; Haiman \& Holder
2003). For example, reionization up to an ionized fraction $x_e=0.3, 0.5.$
and $0.7$ at $z_Q=34, 24,$ and $20$, respectively, with $x_e=const$ down to
$z\sim 7$ and $x_e\sim 1$ at $0<z\la 7$ yields $\tau_e=0.17$ (Ricotti \&
Ostriker 2003).  In these {\it preionization} scenarios, secondary
ionizations are important because the quasar spectrum is significantly
harder and consists primarily of X--rays. Since helium is ignored,
$x_e=x_{H^{+}}$. In \S\ref{sec:discuss:helium} we demonstrate that including
helium does not change the final results by more than $\sim 15\%$.

X--rays may have some advantages in partially reionizing the universe (Oh
2001; Venkatesan, Giroux, \& Shull 2001; Ricotti \& Ostriker 2003).  First,
for low ionization states of a gas ($x_{H^{+}} \lesssim 0.1$), a single X--ray
photon can produce multiple secondary ionizations. Also, the escape
fractions of X--rays from the local high density environments of sources
into the lower density regions, is high: $f_{\rm esc}\sim 1$.  In these low
density regions, the clumping factor $C$ is expected to be $\sim 1$.

This implies that, contrary to the case of full ionization by
UV--photons, which requires $\eta \gtrsim 10$ photons per baryon
(\S\ref{sec:method}), partial ionization to a fraction $x_{H^{+}}$ by
X--ray photons requires only $\eta=x_{H^{+}} (1+N_{\rm rec})/(1+\mathcal{N}_{\rm
sec})$, where $\mathcal{N}_{\rm sec}$ is defined to be the mean number of
secondary ionizations caused by a single photon (given below in
eq. [\ref{eq:ngammasec4}]).
In the preionization models of Ricotti \& Ostriker (2003) a
considerably harder template spectrum is used, with $F_E \propto E^4$
at $E < 100$ eV, and $\propto E^{-1}$ at $E > 100 $ eV, as appropriate
for highly obscured sources.  A physical motivation for the break at
this energy is that the column density of isothermal gas in
hydrostatic equilibrium in a typical virialized halo at $z=10-20$ is $
N_{\rm HI}\sim 10^{20-21} \hs {\rm cm}^{-2}$ (Dijkstra et al. 2004).
Photons with $E \gtrsim 150 \hs \rm{eV}\hs(N_{\rm HI}/10^{20}\hs {\rm
cm}^{-2})^{1/3}$  can escape unimpeded from a halo, whereas the halo is
optically thick to less energetic photons.

In order to repeat the analysis of \S\ref{sec:method} with the
modified spectrum and the secondary ionizations included, we first
compute the mean number of secondary ionizations per photon, $\mathcal{N}_{\rm
sec}$. 
The total number of ionizations per unit comoving volume to go from
electron fraction $x_e$ to $x_e+dx_e$ is:

\be
dn_{\rm Hion}=(1+N_{\rm rec}) \hs dx_e \hs n_{H,0}
\label{eq:ngammasec}
\ee

The number of hydrogen ionizations per comoving volume is given by:

\begin{eqnarray}
\nonumber n_{\rm Hion}=n_{\gamma}\Big{(}1+\frac{1}{n_{\gamma}}
\frac{1+z}{13.6 \hs \rm{eV}} \frac{4\pi}{c}\int_{\frac{13.6}{1+z} \rm
{eV}}^{\frac{E_{\rm{max}}}{1+z} \rm {eV}} dE \hs F_E \times \hs\\
\times \phi[ E(1+z)-13.6 \hs \rm{eV},x_e ]\Big{)}\equiv
n_{\gamma}[1+\langle \mathcal{N}_{\rm sec}\rangle(x_e)]
\label{eq:ngammasec2}
\end{eqnarray}

In this equation, $n_{\gamma}$ is the comoving number density of H--ionizing
photons (eq. \ref{eq:ngamma}). As mentioned above, the more energetic photons
of energy $E$ will produce a secondary photo--electron of energy $E-13.6
\hs \rm{eV}$, which will spend a fraction $\phi(E-13.6 \hs \rm{eV},x_e)$ of
its energy on H--ionizations.  The fitting formula we used for $\phi(E,x_{H^{+}})$ 
in our numerical integration is described in the Appendix. 
Note that we can only write $x_e$ here
because it is equal to $x_{H^{+}}$ (see \S\ref{sec:discuss:helium} for the
effect of helium).  The quantity $\langle \mathcal{N}_{\rm sec}\rangle$ $(x_e)$
denotes the mean number of secondary ionizations per photon over the entire
energy range at electron fraction $x_e$. The above equation simply states
that each H--ionizing photon produces one primary and $\langle \mathcal{N}_{\rm sec}\rangle$ $(x_e)$
secondary ionizations, when the electron fraction in the gas is $x_e$.
Note that the primary ionizations are actually driven by
helium, but as argued in \S\ref{sec:discuss:helium} this only slightly
changes our final results.  

Combining equations (\ref{eq:ngammasec}) and (\ref{eq:ngammasec2}), we
require for the number of ionizing photons per comoving volume

\be dn_{\gamma}(x_e\rightarrow x_e+dx_e)=
\frac{(1+N_{\rm rec}) \hs n_{H,0}\hs dx_e}{1+\langle \mathcal{N}_{\rm sec}\rangle (x_e)}.
\label{eq:ngammasec3}
\ee

The total number of ionizing photons per comoving volume, $n_{\gamma,
\hs \rm{tot}}$, to change the ionized fraction from $0$ to $x_e$ can be
obtained by integrating equation (\ref{eq:ngammasec3}) from $0$ to
$x_e$.  

\be n_{\gamma, \hs \rm{tot}}\equiv\frac{(1+N_{\rm rec}) \hs x_e\hs
n_{H,0}}{1+\mathcal{N}_{\rm sec}},
\label{eq:ngammasec4}
\ee in which the number $\mathcal{N}_{\rm sec}$ is
the mean number of secondary ionizations per photon over the whole
energy and electron fraction range. This is the number we referred to earlier in this section.

To show the relative importance of secondary ionizations, we list the
values of $\mathcal{N}_{\rm sec}$ in Table~\ref{table2}. Note that these numbers
are slightly higher than those in Madau et al. (2003). The main reason
is that our number is mean number of secondary ionizations per photon
over the whole electron fraction range up to $x_e$, whereas their
number is given at the specific value $x_e$.

Given $\mathcal{N}_{\rm sec}$, we set
$\eta=x_e(1+N_{\rm rec})/(1+\mathcal{N}_{\rm sec})$ and calculate the
present-day SXB intensity for the three preionization scenarios with
different $x_e$ and reionization redshift, as considered by Ricotti \&
Ostriker (2003). The fluxes we find are listed in Table~\ref{table2}.
As the table shows, models within which X--rays are assumed to partially ionize the
IGM to $x_e\lesssim 0.5$ do not saturate the mean unaccounted SXB.
The preionization scenario with $x_e=0.7$ has a similar
contribution to the SXB as the models that aim for full reionization and saturates the
maximum unaccounted flux.

As mentioned in \S\ref{sec:method} the results are dependent on the exact
value of $E_{\rm max}$, being the energy below which photons would
typically be absorbed and contribute to H-reionization at any redshift
prior to $z\approx 6$, for a partially ionized IGM. The optical depth of a
partially ionized IGM between $z=z_Q$ and $z=6$ for a photon of energy $E$ emitted at
$z=z_Q$ is, \be \label{eq:emax}\tau(E)=\frac{c}{H_0
\sqrt{\Omega_m}}\int_{6}^{z_Q}
\frac{dz}{(1+z)^{5/2}}\big{[}n_{HI}\sigma_H(E')+
n_{HeI}\sigma_{He}(E')\big{]}, \ee 
where $E'=E(1+z)/(1+z_Q)$ and $n_{HI}
\hs(n_{HeI})$ is the proper number density of neutral hydrogen (helium) atoms.
Although helium has been ignored in this section, it
is included in this formula since it dominates
the contribution to $\tau(E)$ for X--ray photons. We discuss
the implications of helium in more detail in \S\ref{sec:discuss:helium}.
The cross--sections are taken from Verner et al. (1996).

The value of $E_{\rm max}$ was obtained by setting, $\tau(E_{\rm
max})\equiv 1$. There is some arbitrariness in this definition, and to
check how it affects our results we repeated our analysis using $E_{\rm
max,alt}$ defined by $\tau(E_{\rm max,alt})\equiv 0.1$.  This roughly
corresponds to doubling $E_{\rm max}$ and brings down our calculated
contributions to the SXB by $\lesssim 15\%$. Since this choice of $E_{\rm
max,alt}$ is very unrealistic (photons of energy $E_{\rm max,alt}$ are
only absorbed $\sim 10\%$ of the time) we believe that the actual choice of
$E_{\rm max}$ is influencing our results by less than $\sim 15\%$.

Using the model for secondary ionizations, we can calculate their
contribution for the full reionization cases described in
\S\ref{sec:fullion}. We find that for the quasar case
(\S\ref{sec:quasars}), $\mathcal{N}_{\rm sec} \lesssim 0.04$, which is a negligible
contribution. For the miniquasars, with harder spectra, $\mathcal{N}_{\rm sec}
\lesssim 0.10$, which would come down to a $\sim 10\%$ reduction in the
calculated contribution to the SXB.  Since the errors in these
contributions were of the order of $\sim 30\%$, ignoring the contributions
from secondary ionizations is justified.

\section{Discussion}
\label{sec:discuss}

\subsection{Constraints on Spectrum}

\begin{figure}[ht]
\vbox{ \centerline{\epsfig{file=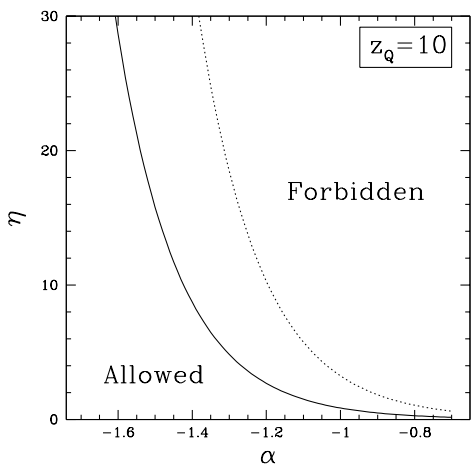,width=8.0truecm}}
\caption{Constraints on the number of ionizing photons per H atom,
$\eta$, (eq. \ref{eq:eta}) and the power--law index for the slope of
the ionizing background, $\alpha$, based on the intensity of the
present--day SXB. The quasars are assumed to form at $z_Q=10$ and have
a power--law spectrum, $F_E \propto E^{-\alpha}$ for $E> 13.6$ eV.
The curves bracket the allowed parameter space for the mean (solid
line) or maximum (dotted) unaccounted flux in the SXB. }
\label{fig:constraints}}
\end{figure}

As emphasized above, the spectral shape of the putative typical
high--$z$ accreting BH is uncertain; the existing templates, motivated by
lower--redshift sources, can be considered merely as guides. In this
section, we adopt a different strategy and ignore any physically
motivated features in the spectrum. Instead, we adopt a simple
power--law shape, $F_E \propto E^{-\alpha}$ for $E> 13.6$ eV.  We then
consider a range of values for $\alpha$, and also vary $\eta$
(eq. \ref{eq:eta}) to calculate the contribution to the SXB.
Figure \ref{fig:constraints} shows which combinations of $\alpha$ and
$\eta$ value (Eq. \ref{eq:eta}) are allowed by the mean and maximum
unaccounted flux in the SXB (see Table \ref{table}).  The solid
(dotted) lines denote those models with combinations of $\alpha$ and
$\eta$ such that the predicted contribution to the SXB is equal to the
mean (maximum) unaccounted flux.  The redshift of the burst of accreting BH formation
in this case is assumed to be $z_Q=10$.

This figure shows that for $\eta=10$, which was used in \S\ref{sec:quasars}
and \S\ref{sec:miniquasars}, any power--law shallower than $\alpha \approx
1.4$ (1.2) will saturate the mean (maximum) unaccounted flux.  This figure
also demonstrates that when $\alpha=1$, the mean (or maximum) unaccounted
flux is saturated when $\eta$ is greater than 1 (or 4).

\subsection{The Influence of Helium}
\label{sec:discuss:helium}
So far our analysis has ignored the presence of helium for simplicity.  In
this section we will demonstrate how the presence of helium reduces the
contribution of accreting black holes to the SXB, albeit at a level
$\lesssim 15\%$.

The photoionization cross--section for neutral (and singly ionized)
helium, $\sigma_{\rm He}$ (and $\sigma_{\rm He^{+}}$) is significantly
larger in the X--rays than that for hydrogen, $\sigma_H$.  For
example, for $E=100$ eV $\sigma_{\rm He}/\sigma_H \sim 20$ and for
greater energies rapidly increases to $\sim 30$ after which this value
stays constant.  Since the number density of helium atoms is $\sim
8\%$ that of hydrogen, photoionization of helium will dominate those
of hydrogen by a factor of $\sim 2-3$. The primary photo--electrons
created will cause secondary ionizations of mainly hydrogen according
to the description in \S\ref{sec:preion}.

For simplicity, let us first neglect recombinations and assume that only
helium is photoionized once, after which the generated photo--electrons
spend $36\%$ of their energy on hydrogen ionizations and $5\%$ on helium
ionizations (Shull \& van Steenberg 1985).  With our assumed spectrum and
taking $z_{Q}=34$ we obtain 8.3 secondary ionizations of hydrogen and 1.1
of helium per photoionization of helium. This implies that for every
ionized helium atom, $8.3/2.1=4$ hydrogen atoms are ionized.  Since the
hydrogen atoms outnumber the helium atoms by a factor of $\sim 12$,
$x_{He^{+}}$ rises faster than $x_{H^{+}}$ by a factor of $\sim 3$ for a neutral
medium.

This ratio is reduced when direct photoionization of hydrogen is taken
into account.  For photons more energetic than 100 eV, roughly 1
hydrogen atom is photoionized per 2 photoionized helium atoms.  This
results in $x_{He^{+}}$ rising faster than $x_{H^{+}}$ by a factor of $\sim
2.5$. The number density of electrons is
$n_e=n_{H^{+}}+n_{He^{+}}+2n_{He^{++}}$. Initially $n_e
=n_{H^{+}}+n_{He^{+}}=n_H(x_{H^{+}}+0.08x_{He^{+}})\sim 1.2n_{H^{+}}$ (since
$x_{He^{+}}/x_{H^{+}}\sim 2.5$ and $x_{He^{++}}=0$).

In fact, the abundance of He$^{++}$ is negligible even when helium
starts to become significantly ionized, because the recombination time
of He$^{++}$ is shorter than that for hydrogen by a factor of $\sim
5$.  Venkatesan, Giroux, \& Shull (2001) included helium in their
models and found that $x_{He^{++}}\sim 0.01x_{He^{+}}$. When He$^{++}$
recombines, a UV--photon will be emitted that can ionize hydrogen and
helium with roughly equal probability, since $\sigma_{\rm He} \sim
14\sigma_H$ for a 55 eV photon.  Similarly He$^{+}$ will recombine
$N_{\rm rec}$ times emitting a UV--photon, which is more likely to
ionize hydrogen than helium by about a factor of 2. This implies that
a fraction $\sim 2N_{\rm rec}/3(1+N_{\rm rec})$ of the photons that
ionized helium will end up ionizing hydrogen. This will bring down
$x_e$ to $\sim \hs 1.15 x_{H^{+}}$. Once He$^{++}$ recombinations become
more abundant, the ratio $x_e/x_{H^{+}}$ will be reduced further, since
the produced UV--photon can ionize hydrogen.  Therefore, the main
effect of recombinations is to make $x_{He^{+}}$ closer to $x_{H^{+}}$ and
to bring $x_e$ closer to $\sim 1.1x_{H^{+}}$ which were the underlying
assumptions in the models of Shull \& van Steenberg (1985).  This
justifies our use of their fitting formula $\phi(E,x_{H^{+}})$.

To estimate how the above effects influence our results in
\S\ref{sec:preion} we replace $x_e$ in equation~(\ref{eq:ngammasec2}) by
$0.87x_e$ (because $x_e\sim1.15 x_{H^{+}}$ as argued above) and also
multiply the number of secondary ionizations by a factor of $1.14$ to
account for secondary ionizations of helium occur at a level of $14\%$
relative to that of hydrogen (Shull \& van Steenberg 1985).  This generally
leads in a reduction of at most $\sim 15\%$ to the contribution to the SXB.
The 'helium corrected' contributions are given in Table~\ref{table2}. Since
the helium correction is small, the main conclusions will not depend on
whether helium is included or not.

The $\lesssim 15\%$ helium correction should also not be taken too
seriously since the method outlined in \S\ref{sec:preion} is highly
simplified.  We did not follow the temperature evolution of the gas, which
is a strong function of the ionized fraction of the gas. For low ionized
fraction $x_{H^{+}}\lesssim 0.1$, $T$ is expected to be much lower than
$10^4$K, whereas it may become much higher once the gas is ionized; the
value of $T$ affects the recombination rates.  Moreover, the incident
spectrum at a certain location will be a function of time and distance to
the nearest source, since an intervening column of neutral hydrogen and
helium (which change as a function of time) will harden the spectrum.  A
proper treatment of this problem requires 3D numerical simulations with
radiative transfer and goes beyond the scope of the present paper.  Keeping
in mind the simplifications inherent in our discussion, the helium
correction is most likely within the uncertainty of the model.

In summary, our estimates of the present--day contribution to the SXB
presented in \S\ref{sec:preion} were slightly too high, because we
ignored helium.  We demonstrated that the amount we overestimated this
contribution is less than a factor of $\sim 1.15$, which is most likely
within model uncertainties.

\subsection{Effects of Continuous (Mini)Quasar Formation}

Our analysis has so far required that all quasars at $z=z_Q$ produce $\eta$
ionizing photons per hydrogen atom, as the condition for achieving complete
reionization. A more realistic model would include a continuous formation
of accreting BHs over an extended redshift interval.  There is no reason to
assume that their formation is stopped once the number of
photons required to ionize the IGM is produced.  Our previous conclusions are therefore
conservative, in the sense that the accreting BHs that form between $z_{\rm
min}\lsim z<z_Q$ would represent an additional contribution to the
unresolved SXB.  Here $z_{\rm min}\sim 4$ is the highest redshift at which
faint quasars would be resolved in existing X-ray surveys (Barger et
al. 2002, 2003); quasars at $z<z_{\rm min}$ are already included in the
resolved fraction of the SXB.

The presence of accreting BHs forming beyond those needed for reionization
will increase the total quasar contribution to the SXB and tighten our
constraints.  Naively, the correction for this effect would increase
with $z_Q$, since the range $z_{\rm min}\lsim z<z_Q$ available for the
continued formation of accreting BHs is longer.  As was shown in
Figure (\ref{fig:saz}) and Table \ref{table2}, the contribution to the
SXB from an ionizing background is higher at lower redshift at a fixed
value of $\eta$.  Modeling the magnitude of the correction due to the
extra accreting BHs is beyond the scope of this paper; we merely emphasize
here that the presence of these extra sources would tighten our
constraints.

\subsection{Known X--ray Point Sources}
\label{sec:discuss:point}

As discussed in \S\ref{sec:sxrb}, a fraction of $88\pm7 \%$ of the SXB
is accounted for by known point sources.  We next discuss whether some
of these point sources may in fact be unidentified high redshift
quasars.  The expected X--ray flux from an accreting BH at redshift
$z_Q$, shining at the Eddington Luminosity, for a black hole of mass
$M_{\rm bh}$ is,
\begin{eqnarray}
\nonumber
S_{\rm QSO}=3.8 \times 10^{-20}\Big{(}\frac{M_{\rm bh}}{10^4 M_{\odot}}\Big{)}\times\\
\times \Big{(} \frac{f_X}{0.03}\Big{)}\Big{[}\frac{10}{(1+z_Q)-1}\Big{]}^{9/4} \hs \hs
\rm{ergs}\hs\rm{sec}^{-1}\hs\rm{cm}^{-2}
\end{eqnarray} 
where $f_X$ is the fraction of the total emitted energy in the
$(0.5-2.0)(1+z_Q)$ keV band.  The value $f_X=0.03$ corresponds to the
template spectrum as given by Sazonov, Ostriker, \& Sunyaev (2004).
The last term is a fit to $d^2_L(10)/d^2_L(z_Q)$, where $d_L(z_Q)$ is
the luminosity distance to $z=z_Q$ (Pen, 1999).  The fit is accurate
to $\sim 4\%$ over the range of $6< z<34$.

For reasonable choices of parameters, we do not expect any of the
known point sources to be the low--mass miniquasars we considered
above. However, massive black holes $M_{\rm bh}\gtrsim 10^8\hs
M_{\odot}$ with harder spectra could have detectable fluxes, and could
be included in the sample of known faint X--ray point sources. The
work by Barger et al. (2002, 2003) suggests that most of the known
point sources in the Chandra Deep Field South (CDF-S) are at redshifts
$z \lesssim 4$. Nevertheless, Koekemoer et al. (2004) find 7 faint
sources in the CDF-S, with counterparts in near IR $JHK$-bands with no
flux in any of the bands up to 8500\AA .  Koekemoer et al. (2004)
speculate that these 7 sources with extreme X--ray/optical flux ratios
(EXOs), could be high--redshift quasars ($6<z<10$).  However, the
total background flux from these seven EXOs is $8.4 \times
10^{-15}~{\rm ergs~s^{-1}~cm^{-2}}$ in the 0.5-8.0 keV band, which
corresponds to an SXB flux of $7.6 \times 10^{-14}~{\rm
ergs~s^{-1}~cm^{-2}~deg^{-2}}$ if such sources populate the whole sky.
Therefore, we conclude that even if all of these sources turn out to
be quasars at $z > 6$, their contribution to the SXB is still $
\lesssim 1\%$ of the total, increasing the allowed SXB from the total
high--redshift accreting BH population by $ \lesssim 25\%$.

\subsection{Reionization Topology and $\eta$}
\label{sec:eta}

We mentioned in \S\ref{sec:method} that the exact value of $\eta$ depends
on the topology of reionization.  We saw that $\eta$ is very different
between a case in which reionization is mediated by UV and X--ray photons.
In the first case, initially $\eta \gtrsim 10$ because the UV--photons
cannot escape the high density regions in which the effective clumping
factor $C$ (eq. \ref{eq:recomb2}) is boosted by small scale density
fluctuations to values $C\gtrsim 10$ (Haiman, Abel, Madau 2001).  After the
filaments are cleared of their neutral hydrogen gas the UV--photons can
escape freely into the low density regions, in which $C\sim 1$. In this
type of ``inside--out'' scenario, the clumping factor is a decreasing
function of cosmic time, and $\eta$ is large ($\gsim 10$).  In the second
case, X--rays are able to escape the filaments unimpeded and immediately
start partially ionizing the low--density voids. Reionization may then be
completed by stars producing UV photons. This scenario does not, in fact,
require hard spectra, as long as the sources send most of their ionizing
radiation directly into voids, rather than ionizing local dense regions.The
clumping factor in this ``outside--in'' case is an increasing function of
cosmic time, and $\eta$ can be low (e.g Miralda--Escud\'e et al. 2000), as
long as most of the mass at $z\sim 7$ is still not highly ionized.  Both
scenarios are theoretically plausible, and depend on the nature and the
location of the first ionizing sources.  Although it is known that in the
local ($z\sim 0$) universe, galaxies lie along dense filaments, the exact
distribution of the high redshift ionizing sources is uncertain and will
influence $\eta$.

\section{Conclusions}
\label{sec:conclusions}

Our main results, obtained in \S~\ref{sec:quasars} and
\S~\ref{sec:miniquasars}, are summarized in Table~\ref{table2}.  The
third column in this table denotes the percentage of the total SXB
contributed by accreting BHs (which we used to refer
to both quasars and their lower--mass ``miniquasar'' counterparts) 
for models in which high redshift accreting BHs
contribute significantly to full or partial reionization.  These
percentages can be compared directly with the percentages given by
M03. The second column gives the total intensity of the quasar
populations, to be compared to the allowed range of the mean or
maximum unresolved SXB flux of $0.35 \times 10^{-12}$ \unitE to $1.23
\times 10^{-12}$ \unitE, respectively.

As Table~\ref{table2} shows, models in which $z>6$ accreting BHs contribute
to reionization overproduce the SXB.  Pre-ionization by miniquasars
requires fewer ionizing photons, both because only a fraction of the H
atoms need to be ionized, and because hard X--rays can produce
multiple secondary ionizations.  We find that models in which X--rays
are assumed to partially ionize the IGM up to $x_e \sim 0.5$ at
$6\la z\la 20$ are still allowed, but could be severely constrained by improved
determinations of the unresolved component of the SXB.

We emphasize that our constraints derive from the total number of ionizing
photons that the population as a whole needs to produce.  Therefore, our
conclusions depend mostly on the assumed spectral shape, and are
independent of the details of the population, such as the luminosity
function and its evolution with redshift.  Future improvements in resolving
the SXB, improving the limits on the unresolved component by a factor of a
few, would place stringent constraints on the contribution of $z\sim 15$
accreting BHs to the scattering optical depth measured by \wmap.

\acknowledgments

This work was supported in part by NASA grant NAG 5-13292, and by NSF
grants AST-0071019, AST-0204514 (for A. L.).

\begin{table*}[ht]
\small
\caption{Unaccounted Soft X-ray Background (0.5-2.0 keV) vs. Quasar/Mini--Quasar Contribution}
\label{table2}
\begin{center}
\begin{tabular}{lccc}
\tablewidth{3in}
Source                   & Flux   & percentage ($\%$) &$\mathcal{N}_{\rm sec}$\\
          & ($10^{-12}$ \unitE)   & of total SXB \\
\hline
\hline
Unaccounted flux (mean)   &  0.35 $\pm$ 0.60 & 4.5 $\pm$ 8.0 &\\
\hline
Unaccounted flux (max)   &  1.23       & 16.3 & \\
\hline
\multicolumn{4}{c}{Quasars \hs \S \ref{sec:quasars}}\\
\multicolumn{4}{c}{Units of Flux here $(\eta/10)\hs10^{-12}$ \unitE}\\
\hline
$z_{\rm Q}=6$&  2.17       & 29 & - \\ 
$z_{\rm Q}=10$&  1.84       & 24 & - \\ 
$z_{\rm Q}=20$&  1.37       & 18 & - \\
\hline
\multicolumn{4}{c}{Mini--Quasars \hs \S \ref{sec:miniquasars} }\\
\hline
$z_{\rm Q}=6$&  $3.3\pm 1.1$       &  $44\pm14$ & - \\ 
$z_{\rm Q}=10$&  $2.3\pm0.7$       &  $30\pm11$ & - \\ 
$z_{\rm Q}=20$&  $1.4\pm0.4$       &  $19\pm 5$ & - \\ 
\hline
\multicolumn{4}{c}{Pre-ionization Scenarios \hs \S\ref{sec:preion} }\\
\multicolumn{4}{c}{Units of Flux here  $\big{(}\frac{\eta \hs (1+\mathcal{N}_{\rm sec})}{(N_{\rm rec}+1) \hs x_e}\big{)}\hs10^{-12}$ \unitE}\\
\hline
$x_e$=0.3, $z_Q=34$ &0.22     & 3 & 2.44\\
$x_e$=0.5, $z_Q=24$ &0.72    & 10 & 1.50\\
$x_e$=0.7, $z_Q=20$ &1.5     & 20 & 0.97\\
\hline
\multicolumn{4}{c}{Helium correction applied to models above (see \S\ref{sec:discuss:helium}).}\\
\hline
$x_e$=0.3, $z_Q=34$ & 0.19    & 3 & 3.03\\
$x_e$=0.5, $z_Q=24$ & 0.61    & 8 & 1.94\\
$x_e$=0.7, $z_Q=20$ &  1.3   & 17 & 1.30\\
\hline
\end{tabular}\\[12pt]
\end{center}
\end{table*}

\section*{Appendix}
We used a fitting formula for the fraction $\phi(x,E)$ of the energy
of a fast photo--electron used for H--ionizations, which is a modified
version of the formula provided by Shull \& van Steenberg (1985).
Their work focusses mostly on energies $E\gg0.1$ keV, in which
$\phi(x,E)$ is independent of energy. The energy dependence at lower
energies, $0.05 < E \lesssim 0.5$ keV, can be seen most clearly in their Figure 3, to which
the formula below provides an accurate fit:

\begin{eqnarray}
\nonumber
\phi(x,E)= 0.39\left[ 1-x^{0.4092 \hs a(x,E)} \right]^{1.7592}\\
a(x,E)=\frac{2}{\pi}\arctan\Big{[}\Big{(}\frac{E}{0.12 \hs \rm{keV}}
\Big{)}\Big{(}\frac{0.03}{x^{1.5}}+1 \Big{)}^{0.25}\Big{]}
\label{eq:phi}
\end{eqnarray}\\

At high energies ($E\gtrsim0.5$ keV) this approaches the fitting formula provided by
Shull \& van Steenberg (1985).
Note that the $E=28$ eV curve is fitted less accurately for $x>0.2$. The
photons in this energy range however, are not affecting our analysis in \S\ref{sec:preion}.

\end{document}